\documentclass[12pt]{article}
\usepackage{axodraw,bbold}

\catcode`@=12
\begin{document}
\thispagestyle{empty}
\begin{flushright} 
UCRHEP-T448\\ 
February 2008\
\end{flushright}
\vspace{0.5in}
\begin{center}
{\LARGE	\bf Utility of a Special Second Scalar Doublet \\}
\vspace{1.5in}
{\bf Ernest Ma\\}
\vspace{0.2in}
{\sl Department of Physics and Astronomy, University of California,\\}
\vspace{0.1in}
{\sl Riverside, California 92521, USA\\}
\vspace{1.5in}
\end{center}

\begin{abstract}\
This Brief Review deals with the recent resurgence of interest in adding 
a second scalar doublet $(\eta^+,\eta^0)$ to the Standard Model of particle 
interactions. In most studies, it is taken for granted that $\eta^0$ should 
have a nonzero vacuum expectation value, even if it may be very small.  
What if there is an exactly conserved symmetry which ensures 
$\langle \eta^0 \rangle = 0$?  The phenomenological ramifications of this 
idea include dark matter, radiative neutrino mass, leptogenesis, and grand 
unification.
\end{abstract}

\newpage
\baselineskip 24pt

The Minimal Standard Model (SM) of particle interactions has only one 
scalar doublet $\Phi = (\phi^+,\phi^0)$.  As $\phi^0$ acquires a nonzero 
vacuum expectation value (vev) $v$, the famous Higgs mechanism allows the 
$W^\pm$ and $Z^0$ gauge bosons to become massive.  At the same time, of 
the original four degrees of freedom in $\Phi$, only one [i.e. the Higgs 
boson $h = \sqrt{2} (Re \phi^0 - v)$] remains.  What happens if more scalar 
doublets are added?  One possibility is to allow for new sources of 
CP nonconservation, as was pointed out a long time ago \cite{l73,w76}. 
Another is to allow the SM to be extended to include supersymmetry, as 
is well-known.  In these and other investigations of the two (or more) 
scalar doublets \cite{koo07}, the usual implicit assumption is that 
both scalar doublets have vev's.  This is necessary in the Minimal 
Supersymmetric Standard Model (MSSM) because one scalar doublet couples 
only to $u$ quarks, and the other only to $d$ quarks and charged leptons.  
However, in the context of the SM alone, a second scalar doublet, call it 
$\eta = (\eta^+,\eta^0)$, is not required to have any vev, in which case an 
exactly conserved $Z_2$ discrete symmetry may be defined, as pointed 
out many years ago \cite{dm78}, implying a stable particle.  This idea 
has been revived recently, together with some new developments.

Following Ref.~\cite{dm78}, consider the simplest possible discrete 
symmetry, i.e. $Z_2$, under which $\eta$ is odd and all SM particles are 
even, it was pointed out first in Ref.~\cite{m06-1} that either $H^0 = 
\sqrt{2}(Re \eta^0)$ or $A^0 = \sqrt{2}(Im \eta^0)$ may be considered as 
a dark-matter candidate.  For this to work, there has to be a splitting 
in mass between the two, otherwise they become just one particle exactly 
like the scalar neutrino of the MSSM, which can interact with nuclei 
through the $Z^0$ boson, having a cross section some eight orders of 
magnitude larger than the present experimental limit.  If the mass 
splitting is larger than about 1 MeV, $Z^0$ exchange is forbidden by 
kinematics in these underground direct-search experiments based on the 
elastic scattering of dark matter with nuclei.  The source of this mass 
splitting is the allowed term $(\lambda_5/2)(\Phi^\dagger \eta)^2 + H.c.$, 
where $\Phi=(\phi^+,\phi^0)$ is the SM Higgs doublet.  This minimal version 
of dark matter has also been proposed \cite{bhr06}, starting with a 
different perspective, and studied seriously \cite{lnot07}.  Its 
astrophysical \cite{glbe07} and collider \cite{cmr07} signatures have 
also been investigated.  The mass of the dark-matter candidate is likely 
to be between 45 and 75 GeV, with detection in direct-search experiments 
at the level of two orders of magnitude below present bounds 
\cite{bhr06,lnot07}.  Its production at the Large Hadron Collider (LHC) 
is from $pp \to Z^0 \to A^0 H^0$ and may be detected through the decay 
$A^0 \to H^0 l^+ l^-$ \cite{bhr06,cmr07}.

With the assumed existence of the special scalar doublet $\eta$, which 
may be called the \emph{dark} scalar doublet (which is more suitable than 
the name ``inert Higgs doublet'' because it is neither inert since it 
has electroweak interactions, nor a ``Higgs'' doublet since it has no vev), 
other particles which are odd under $Z_2$ may be contemplated.  Indeed, 
in that first paper \cite{m06-1} already mentioned, three heavy neutral 
Majorana fermions $N_i$ were proposed which are also odd under $Z_2$. 
This means that the Yukawa terms $(\nu_i \phi^0 - l_i \phi^+)N_j$ are 
forbidden, so that $N_i$ are not Dirac mass partners of $\nu_i$.  On the 
other hand, the Yukawa terms $(\nu_i \eta^0 - l_i \eta^+)N_j$ are allowed, 
so that one-loop radiative Majorana seesaw masses for $\nu_i$ may be 
generated, as shown in Fig.~1.

\begin{figure}[htb]
\begin{center}
\begin{picture}(360,120)(0,0)
\ArrowLine(90,10)(130,10)
\ArrowLine(180,10)(130,10)
\ArrowLine(180,10)(230,10)
\ArrowLine(270,10)(230,10)
\DashArrowLine(155,85)(180,60)3
\DashArrowLine(205,85)(180,60)3
\DashArrowArc(180,10)(50,90,180)3
\DashArrowArcn(180,10)(50,90,0)3

\Text(110,0)[]{$\nu_i$}
\Text(250,0)[]{$\nu_j$}
\Text(180,0)[]{$N_k$}
\Text(135,50)[]{$\eta^0$}
\Text(230,50)[]{$\eta^0$}
\Text(150,90)[]{$\phi^{0}$}
\Text(217,90)[]{$\phi^{0}$}

\end{picture}
\end{center}
\caption{One-loop generation of neutrino mass.}
\end{figure}
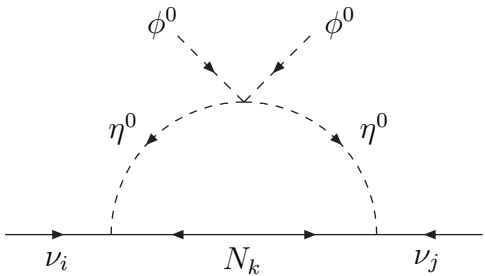

In the canonical seesaw mechanism, doublet neutrinos acquire mass through 
mixing with heavy singlet neutral fermions (often called right-handed 
neutrinos).  Here there is no mixing at all.  Radiative masses appear 
from electroweak symmetry breaking, i.e. $\langle \phi^0 \rangle = v$, 
which is also the source of $A^0-H^0$ mass splitting.  In other words, 
the same mechanism which allows $H^0$ to be a suitable dark-matter 
candidate also allows $\nu_i$ to acquire nonzero radiative seesaw masses. 
Specifically, the diagram of Fig.~1 is exactly calculable from the 
exchange of $H^0$ and $A^0$, i.e.
\begin{equation}
({\cal M}_\nu)_{ij} = \sum_k {h_{ik} h_{jk} M_k \over 16 \pi^2} \left[ 
{m_H^2 \over m_H^2-M_k^2} \ln {m_H^2 \over M_k^2} - {m_A^2 \over m_A^2-M_k^2} 
\ln {m_A^2 \over M_k^2} \right],
\end{equation}
where $M_k$ is the Majorana mass of $N_k$.  Using $m_H^2-m_A^2 = 2 \lambda_5 
v^2$ and $m_0^2 \equiv (m_H^2 + m_A^2)/2$, and assuming $m_0^2 << M_k^2$, then
\begin{equation}
({\cal M}_\nu)_{ij} = {\lambda_5 v^2 \over 8 \pi^2} \sum_k {h_{ik} h_{jk} \over 
M_k} \left[ \ln {M_k^2 \over m_0^2} - 1 \right].
\end{equation}
This formula shows that smaller values of $M_k$ than those of the canonical 
seesaw mechanism may be used for generating the same neutrino masses. 

In canonical leptogenesis \cite{bpy05}, the decay $N \to \phi^\pm l^\mp$ 
generates a lepton asymmetry in the early Universe which gets converted 
into a baryon asymmetry through sphalerons.  Here the decay is instead 
$N \to \eta^{\pm} l^\mp$, which connects the existence of dark matter 
to the baryon asymmetry of the Universe \cite{m06-2}.  In the minimal 
supersymmetric version \cite{m06-3} of this model, there is also a bonus. 
Because of the radiative suppression of Eq.~(2), the mass of the lightest 
$N_i$ may now be safely below the Davidson-Ibarra bound \cite{di02} of 
about $10^9$ GeV.

Another very important consequence is the emergence of two or more types 
of dark matter.  After all, there is no fundamental principle which requires 
that there is only one type of dark matter \cite{bfs04,hln07}.  A generic 
discussion of this possibility has recently appeared \cite{cmwy07}.

Suppose the lightest particle odd under $Z_2$ is $N_k$, then it may also 
be a dark-matter candidate \cite{kms06}, but severe constraints from 
lepton flavor violating processes such as $\mu \to e \gamma$ become 
important because $N_k$ annihilates only through its Yukawa couplings to 
leptons and its proper relic abundance requires those to be large. 
One way to escape such constraints is to allow $N_k$ to have additional 
interactions from an extra U(1) gauge group \cite{ks06} or an extra 
scalar singlet \cite{bm07}.

Since $\mu \to e \gamma$ is automatically allowed in this class of models, 
the muon anomalous magnetic moment must also have a contribution.  To 
suppress the former and to enhance the latter, a variant of the proposed 
mechansim of radiative neutrino mass has also been proposed \cite{hkmr07}. 
Here lepton number is considered as a global $U(1)_L$ symmetry with heavy 
neutral fermion singlets $N_i$ and $N^c_i$ transforming as $1$ and $-1$ 
respectively.  They then have allowed Dirac masses, but they are also 
odd under $Z_2$.  Together with the usual $\eta$ doublet and a new scalar 
charged singlet $\chi^-$, the Yukawa terms $(\nu_i \eta^0 - l_i \eta^+)N^c_j$ 
and $l^c_i \chi^- N_j$ are allowed, resulting in enhanced contributions 
to the muon anomalous magnetic moment.  As for neutrino mass, it again 
occurs in one loop, but only if $U(1)_L$ is broken down to $(-)^L$, which 
may be accomplished by the small explicit soft terms $N_i N_j$ and 
$N^c_i N^c_j$.  Hence the formula for radiative neutrino mass has one more 
suppression, which argues for the Dirac masses of $(N,N^c)$ to be of order 
TeV.  Leptogenesis may also be implemented by the decay of the lightest 
such pair.

Given the structure of the minimal dark scalar doublet model or its 
supersymmetric extension, the next question to ask is whether it has 
a natural grand unification.  There have been two developments.  
One is to consider its supersymmetric SU(5) completion \cite{m08-1}.  
This means adding the superfields
\begin{eqnarray}
&& \underline{5} = h (3,1,-1/3) + (\eta^+_2,\eta^0_2) (1,2,1/2), \\ 
&& \underline{5}^* = h^c (3^*,1,1/3) + (\eta^0_1,\eta^-_1) (1,2,-1/2),
\end{eqnarray}
both of which are odd under $Z_2$.  Conventionally, the existence of $h(h^c)$ 
in the $\underline{5}(\underline{5}^*)$ representations of SU(5) is 
considered dangerous because it would mediate rapid proton decay.  
However, the new $Z_2$ symmetry used here for dark matter also serves the 
purpose of conserving baryon number and preventing proton decay.  The 
signature of this model is the decay $h \to d e^- \eta^+_2$ or 
$d e^+ \eta^-_2$.  The production of $h \bar{h}$ will thus result in 
same-sign dileptons plus quark jets plus missing energy.

Another possible embedding of the dark scalar doublet model is into 
the supersymmetric $E_6/U(1)_N$ model \cite{m96}.  There are now three 
\underline{27} representations of superfields, and two $Z_2$ discrete 
symmetries are imposed \cite{ms07}, as shown below.

\begin{table}[htb]
\caption{Particle content of \underline{27} of $E_6$ under $SU(3)_C \times 
SU(2)_L \times U(1)_Y$ and $U(1)_N$.}
\begin{center}
\begin{tabular}{|c|c|c|}
\hline 
Superfield & $SU(3)_C \times SU(2)_L \times U(1)_Y$ & $U(1)_N$ \\ 
\hline
$Q = (u,d)$ & (3,2,1/6) & 1 \\
$u^c$ & $(3^*,1,-2/3)$ & 1 \\ 
$e^c$ & (1,1,1) & 1 \\
\hline
$d^c$ & $(3^*,1,1/3)$ & 2 \\ 
$L = (\nu,e)$ & $(1,2,-1/2)$ & 2 \\ 
\hline
$h$ & $(3,1,-1/3)$ & $-2$ \\ 
$\bar{E} = (E^c,N^c_E)$ & $(1,2,1/2)$ & $-2$ \\ 
\hline
$h^c$ & $(3^*,1,1/3)$ & $-3$ \\ 
$E = (\nu_E,E)$ & $(1,2,-1/2)$ & $-3$ \\ 
\hline
$S$ & $(1,1,0)$ & 5 \\
\hline
$N^c$ & (1,1,0) & 0 \\ 
\hline
\end{tabular}
\end{center}
\end{table}

\begin{table}[htb]
\caption{Particle content of \underline{27} of $E_6$ under $M$ parity  
and $N$ parity.}
\begin{center}
\begin{tabular}{|c|c|c|}
\hline 
Superfield & $M$ & $N$ \\ 
\hline
$Q,u^c,d^c$ & + & + \\
$L,e^c$ & $-$ & + \\ 
$h,h^c$ & $-$ & + \\
$E_1,\bar{E}_1,S_1$ & $+$ & + \\ 
$E_{2,3},\bar{E}_{2,3},S_{2,3}$ & $+$ & $-$ \\ 
$N^c$ & $-$ & $-$ \\ 
\hline
\end{tabular}
\end{center}
\end{table}

Since the $\lambda_5$ term of the SM is not available in supersymmetry, the 
equivalent $A^0-H^0$ mass splitting is now achieved in one loop, from 
the effective $[(\tilde{N}^c_E)_1^\dagger (\tilde{N}^c_E)_{2,3}]^2$ term after 
supersymmetry breaking.  Neutrino masses are then obtained in two loops. 
It may be noted that two-loop neutrino masses are also naturally 
obtained in a model of $Z_3$ dark matter \cite{m07-1}. Another variant of 
the $E_6/U(1)_N$ model has also been proposed \cite{m07-2} with the 
multiplicative conservation of baryon number.

In all of the above models of a special second scalar doublet, the $Z_2$ 
discrete symmetry is imposed by hand.  Is it possible to obtain it from 
a gauge symmetry?  The answer is yes, as shown in a recent explicit example 
\cite{m08-2}.  The idea is to extend the MSSM with a new $U(1)_X$ gauge 
group, so that the usual $R$ parity is automatic (i.e. not imposed by hand 
as in the MSSM), and the new $Z_2$ is the remnant of $U(1)_X$ breaking.  
This particular realization requires the addition of new superfields as 
shown below.  Under $U(1)_X$, the MSSM superfields $(u,d),(\nu,e)$ are 
trivial; $u^c,(\phi_1^0,\phi_1^-)$ transform as $n_2$, and $d^c,e^c,
(\phi_2^+,\phi_2^0)$ as $-n_2$. For $U(1)_X$ to be anomaly-free, eleven 
copies of $N_i$ are required.

\begin{table}[htb]
\caption{New particle content of $U(1)_X$ model.}
\begin{center}
\begin{tabular}{|c|c|c|}
\hline 
Superfield & $SU(3)_C \times SU(2)_L \times U(1)_Y$ & $U(1)_X$ \\ 
\hline
$\eta_1 \equiv (\eta^0_1,\eta^-_1)$ & $(1,2,-1/2)$ & $n_2/4$ \\ 
$\eta_2 \equiv (\eta^+_2,\eta^0_2)$ & $(1,2,1/2)$ & $-n_2/4$ \\ 
\hline
$N_i$ & $(1,1,0)$ & $n_2/4$ \\
$\chi$ & $(1,1,0)$ & $-3n_2/4$ \\
\hline
$S$ & $(1,1,0)$ & $-n_2/2$ \\
$\zeta$ & $(1,1,0)$ & $3n_2/2$ \\
\hline
\end{tabular}
\end{center}
\end{table}

In summary, the utility of a special second scalar doublet $(\eta^+,\eta^0)$ 
with $\langle \eta^0 \rangle = 0$ has been discussed. It points to an 
exactly conserved $Z_2$ discrete symmetry, which may be important for 
understanding the interconnectedness of dark matter, radiative neutrino 
mass, leptogenesis, and grand unification.  It predicts new particles 
at or below the TeV energy scale which may be verifiable at the 
forthcoming Large Hadron Collider (LHC) at CERN.

This work was supported in part by the U.~S.~Department of Energy under Grant 
No.~DE-FG03-94ER40837.

\bibliographystyle{unsrt}

\begin{thebibliography}{99}
\bibitem{l73} T. D. Lee, Phys. Rev. {\bf D8}, 1226 (1973).
\bibitem{w76} S. Weinberg, Phys. Rev. Lett. {\bf 37}, 657 (1976).
\bibitem{koo07} For a recent study, see for example A. W. El Kaffas, 
P. Osland, and O. M. Ogreid, Phys. Rev. {\bf D76}, 095001 (2007).
\bibitem{dm78} N. G. Deshpande and E. Ma, Phys. Rev. {\bf D18}, 2574 (1978).
\bibitem{m06-1} E. Ma, Phys. Rev. {\bf D73}, 077301 (2006).
\bibitem{bhr06} R. Barbieri, L. J. Hall, and V. S. Rychkov, Phys. Rev. 
{\bf D74}, 015007 (2006).
\bibitem{lnot07} L. Lopez Honorez, E. Nezri, J. F. Oliver, and M. H. G. 
Tytgat, JCAP {\bf 02}, 028 (2007).
\bibitem{glbe07} M. Gustafsson, E. Lundstrom, L. Bergstrom, and J. Edsjo, 
Phys. Rev. Lett. {\bf 99}, 041301 (2007).
\bibitem{cmr07} Q.-H. Cao, E. Ma, and G. Rajasekaran, Phys. Rev. {\bf D76}, 
095011 (2007).
\bibitem{bpy05} For a recent review, see for example W. Buchmuller, R. D. 
Peccei, and T. Yanagida, Annu. Rev. Nucl. Part. Sci. {\bf 55}, 311 (2005).
\bibitem{m06-2} E. Ma, Mod. Phys. Lett. {\bf A21}, 1777 (2006).
\bibitem{m06-3} E. Ma, Annales de la Fondation de Broglie {\bf 31}, 
85 (2006) [hep-ph/0607142].
\bibitem{di02} S. Davidson and A. Ibarra, Phys. Lett. {\bf B535}, 25 (2002).
\bibitem{bfs04} C. Boehm, P. Fayet, and J. Silk, Phys. Rev. {\bf D69}, 
101302 (2004).
\bibitem{hln07} T. Hur, H. S. Lee, and S. Nasri, arXiv:0710.2653 [hep-ph].
\bibitem{cmwy07} Q.-H. Cao, E. Ma, J. Wudka, and C.-P. Yuan, 
arXiv:0711.3881 [hep-ph].
\bibitem{kms06} J. Kubo, E. Ma, and D. Suematsu, Phys. Lett. {\bf B642}, 
18 (2006).
\bibitem{ks06} J. Kubo and D. Suematsu, Phys. Lett. {\bf B643}, 336 (2006).
\bibitem{bm07} K. S. Babu and E. Ma, arXiv:0708.3790 [hep-ph].
\bibitem{hkmr07} T. Hambye, K. Kannike, E. Ma, and M. Raidal, Phys. Rev. 
{\bf D75}, 095003 (2007).
\bibitem{m08-1} E. Ma, Phys. Lett. {\bf B659}, 885 (2008).
\bibitem{m96} E. Ma, Phys. Lett. {\bf B380}, 286 (1996).
\bibitem{ms07} E. Ma and U. Sarkar, Phys. Lett. {\bf B653}, 288 (2007).
\bibitem{m07-1} E. Ma, arXiv:0708.3371 [hep-ph].
\bibitem{m07-2} E. Ma, arXiv:0710.1102 [hep-ph].
\bibitem{m08-2} E. Ma, arXiv:0801.2545 [hep-ph].

\end{thebibliography}

\end{document}